\documentclass[twoside,twocolumn,9pt]{article}
\usepackage{extsizes}
\usepackage[super,sort&compress,comma]{natbib} 
\usepackage[version=3]{mhchem}
\usepackage[left=1.5cm, right=1.5cm, top=1.785cm, bottom=2.0cm]{geometry}
\usepackage{balance}
\usepackage{mathptmx}
\usepackage{sectsty}
\usepackage{graphicx} 
\usepackage{lastpage}
\usepackage[format=plain,justification=justified,singlelinecheck=false,font={stretch=1.125,small,sf},labelfont=bf,labelsep=space]{caption}
\usepackage{float}
\usepackage{fancyhdr}
\usepackage{fnpos}
\usepackage[english]{babel}
\addto{\captionsenglish}{%
  
}
\usepackage{array}
\usepackage{droidsans}
\usepackage{charter}
\usepackage[T1]{fontenc}
\usepackage[usenames,dvipsnames]{xcolor}
\usepackage{setspace}
\usepackage[compact]{titlesec}
\usepackage[hidelinks]{hyperref}

\usepackage{epstopdf}

\definecolor{cream}{RGB}{222,217,201}

\begin{document}

\pagestyle{fancy}
\thispagestyle{plain}
\fancypagestyle{plain}{
\renewcommand{\headrulewidth}{0pt}
}

\makeFNbottom
\makeatletter
\renewcommand\LARGE{\@setfontsize\LARGE{15pt}{17}}
\renewcommand\Large{\@setfontsize\Large{12pt}{14}}
\renewcommand\large{\@setfontsize\large{10pt}{12}}
\renewcommand\footnotesize{\@setfontsize\footnotesize{7pt}{10}}
\makeatother

\renewcommand{\thefootnote}{\fnsymbol{footnote}}
\renewcommand\footnoterule{\vspace*{1pt}%
\color{cream}\hrule width 3.5in height 0.4pt \color{black}\vspace*{5pt}} 
\setcounter{secnumdepth}{5}

\makeatletter 
\renewcommand\@biblabel[1]{#1}            
\renewcommand\@makefntext[1]%
{\noindent\makebox[0pt][r]{\@thefnmark\,}#1}
\makeatother 
\renewcommand{\figurename}{\small{Fig.}~}
\sectionfont{\sffamily\Large}
\subsectionfont{\normalsize}
\subsubsectionfont{\bf}
\setstretch{1.125} 
\setlength{\skip\footins}{0.8cm}
\setlength{\footnotesep}{0.25cm}
\setlength{\jot}{10pt}
\titlespacing*{\section}{0pt}{4pt}{4pt}
\titlespacing*{\subsection}{0pt}{15pt}{1pt}

\fancyfoot{}
\fancyfoot[LO,RE]{\vspace{-7.1pt}\includegraphics[height=9pt]{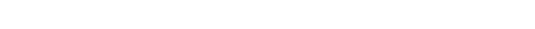}}
\fancyfoot[CO]{\vspace{-7.1pt}\hspace{13.2cm}\includegraphics{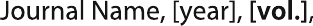}}
\fancyfoot[CE]{\vspace{-7.2pt}\hspace{-14.2cm}\includegraphics{RF}}
\fancyfoot[RO]{\footnotesize{\sffamily{1--\pageref{LastPage} ~\textbar  \hspace{2pt}\thepage}}}
\fancyfoot[LE]{\footnotesize{\sffamily{\thepage~\textbar\hspace{3.45cm} 1--\pageref{LastPage}}}}
\fancyhead{}
\renewcommand{\headrulewidth}{0pt} 
\renewcommand{\footrulewidth}{0pt}
\setlength{\arrayrulewidth}{1pt}
\setlength{\columnsep}{6.5mm}
\setlength\bibsep{1pt}

\makeatletter 
\newlength{\figrulesep} 
\setlength{\figrulesep}{0.5\textfloatsep} 

\newcommand{\topfigrule}{\vspace*{-1pt}%
\noindent{\color{cream}\rule[-\figrulesep]{\columnwidth}{1.5pt}} }

\newcommand{\botfigrule}{\vspace*{-2pt}%
\noindent{\color{cream}\rule[\figrulesep]{\columnwidth}{1.5pt}} }

\newcommand{\dblfigrule}{\vspace*{-1pt}%
\noindent{\color{cream}\rule[-\figrulesep]{\textwidth}{1.5pt}} }

\makeatother

\twocolumn[
  \begin{@twocolumnfalse}
{\includegraphics[height=30pt]{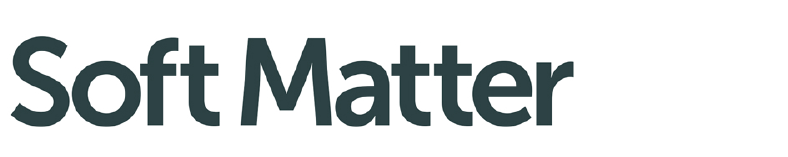}\hfill\raisebox{0pt}[0pt][0pt]{\includegraphics[height=55pt]{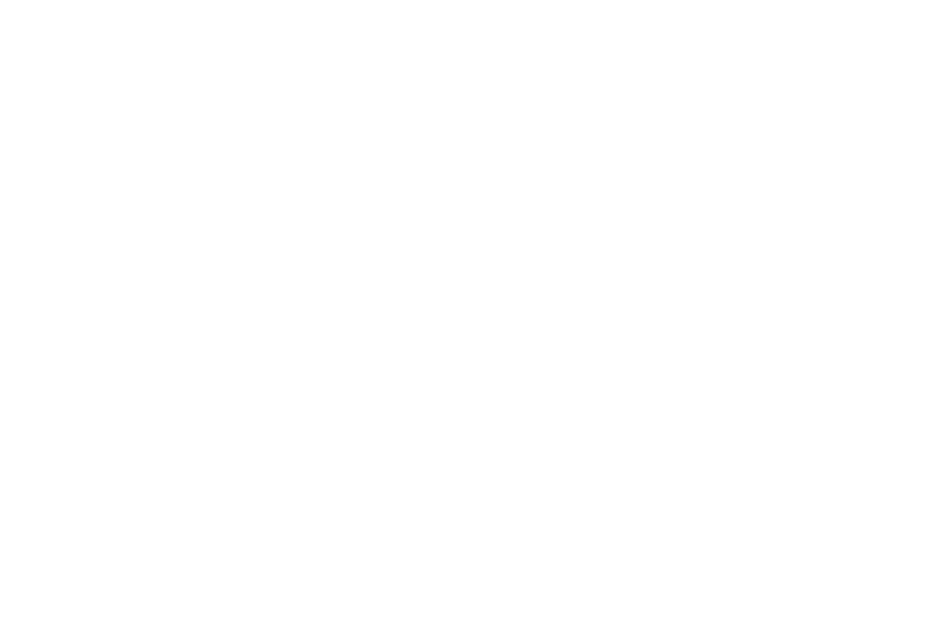}}\\[1ex]
\includegraphics[width=18.5cm]{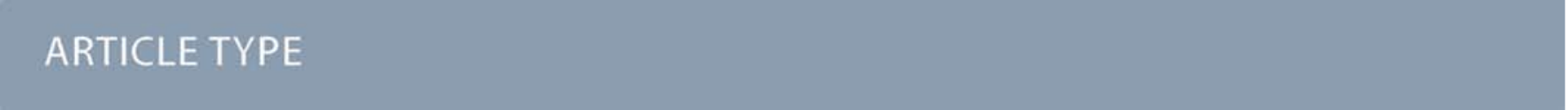}}\par
\vspace{1em}
\sffamily
\begin{tabular}{m{4.5cm} p{13.5cm} }

\includegraphics{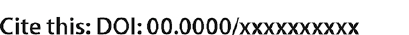} & \noindent\LARGE{\textbf{Collapse of a hemicatenoid bounded by a solid wall: instability and dynamics 
driven by surface Plateau border friction$^\dag$}} \\
\vspace{0.3cm} & \vspace{0.3cm} \\

 & \noindent\large{Christophe Raufaste,\textit{$^{a}$} Simon Cox,\textit{$^{b}$} 
 Raymond E. Goldstein,\textit{$^{c}$} and Adriana I. Pesci\textit{$^{d}$}} \\

\includegraphics{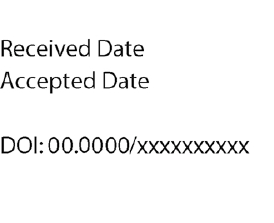} & \noindent\normalsize{
The collapse of a catenoidal soap film when the rings supporting it are moved beyond a critical separation
is a classic problem in interface motion in which there is a balance between surface tension and 
the inertia of the surrounding air, with film viscosity playing only a minor role.  Recently 
[Goldstein, {\it et al.}, {\it Phys. Rev. E}, 2021, 104, 035105], we introduced
a variant of this problem in which the catenoid is bisected by a glass plate located in a 
plane of symmetry perpendicular to the rings, producing two identical hemicatenoids, each with a
surface Plateau border (SPB) on the glass plate.  Beyond the critical ring separation, the hemicatenoids collapse
in a manner qualitatively similar to the bulk problem, but their motion is governed by the
frictional forces arising from viscous dissipation in the SPBs.  
Here we present numerical studies of a model that includes classical friction laws for SPB motion on wet
surfaces and show consistency with our experimental measurements of the temporal evolution of this process.  
This study can help explain the fragmentation of bubbles inside very confined geometries 
such as porous materials or microfluidic devices.} \\

\end{tabular}
\end{@twocolumnfalse} \vspace{0.6cm}
]

\renewcommand*\rmdefault{bch}\normalfont\upshape
\rmfamily
\section*{}
\vspace{-1cm}


\footnotetext{\textit{$^{a}$~Universit{\'e} C{\^o}te d'Azur, CNRS, Institut de Physique de Nice (INPHYNI), 06100 Nice, France
and Institut Universitaire de France (IUF), 75005 Paris, France. E-mail: Christophe.Raufaste@univ-cotedazur.fr}}
\footnotetext{\textit{$^{b}$~Department of Mathematics, Aberystwyth University, Aberystwyth, Ceredigion, SY23 3BZ, United Kingdom. E-mail: simon.cox@aber.ac.uk. }}
\footnotetext{\textit{$^{c}$~Department of Applied Mathematics and Theoretical Physics, Centre for Mathematical Sciences,
University of Cambridge, Wilberforce Road, Cambridge CB3 0WA, United Kingdom. E-mail: R.E.Goldstein@damtp.cam.ac.uk}}
\footnotetext{\textit{$^{d}$~Department of Applied Mathematics and Theoretical Physics, Centre for Mathematical Sciences,
University of Cambridge, Wilberforce Road, Cambridge CB3 0WA, United Kingdom. E-mail: A.I.Pesci@damtp.cam.ac.uk}}
\footnotetext{\dag~Electronic Supplementary Information (ESI) available: Videos from experiments and 
simulations. See DOI: 10.1039/cXsm00000x/}




\section{Introduction}

\begin{figure}[t]
  \centering
  \includegraphics[clip=true,width=0.9\columnwidth]{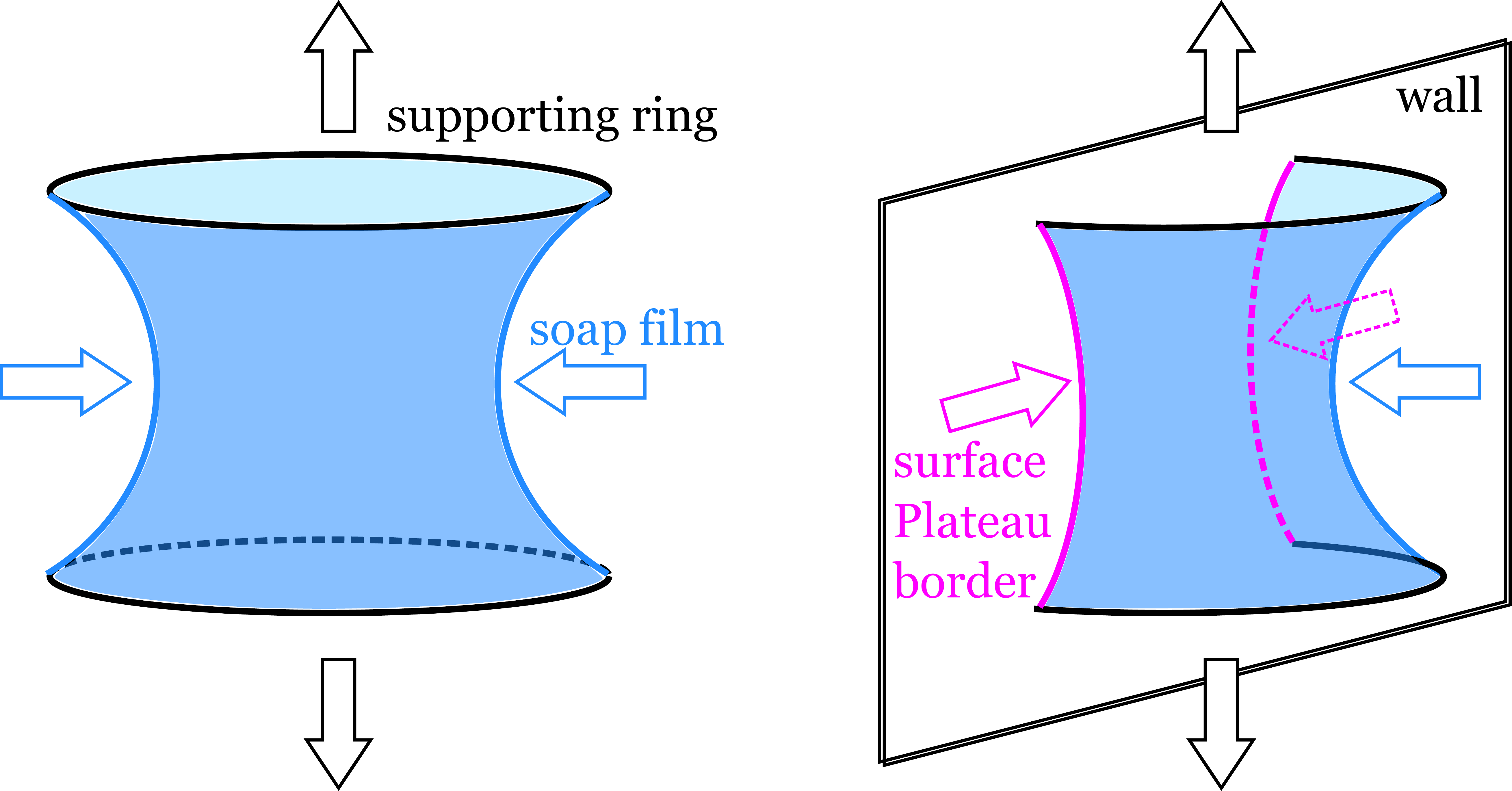}
  \caption{The full catenoid (left) vs the hemicatenoid (right). Stretching the supporting rings of 
each catenoid vertically (black arrows) leads to an instability in which the neck collapses inward.
For the hemicatenoid, that neck narrowing involves inwardly moving surface Plateau borders.}
  \label{fig1}
\end{figure}

Free-boundary problems involving the collapse of unstable minimal surfaces generally fall
into two main categories: either the ultimate singularity and surface reconnection occur in
the bulk, far from any supporting boundaries, or they occur at the boundary itself.  A bulk singularity 
is found in the classical problem of a collapsing catenoid supported by two rings, which was
first studied by Plateau \cite{Plateau1873} and Maxwell \cite{Maxwell} in the XIX\textsuperscript{th} century,
whereas boundary singularities were only studied recently in the context of collapsing 
soap films with more exotic topologies \cite{twist,boundary,systolic}.  A typical example of a 
boundary singularity
is found in the collapse of a M{\"o}bius strip soap film \cite{twist}, 
first studied by Courant \cite{Courant}.

\begin{figure}[t]
  \centering
  \includegraphics[clip=true,width=0.70\columnwidth]{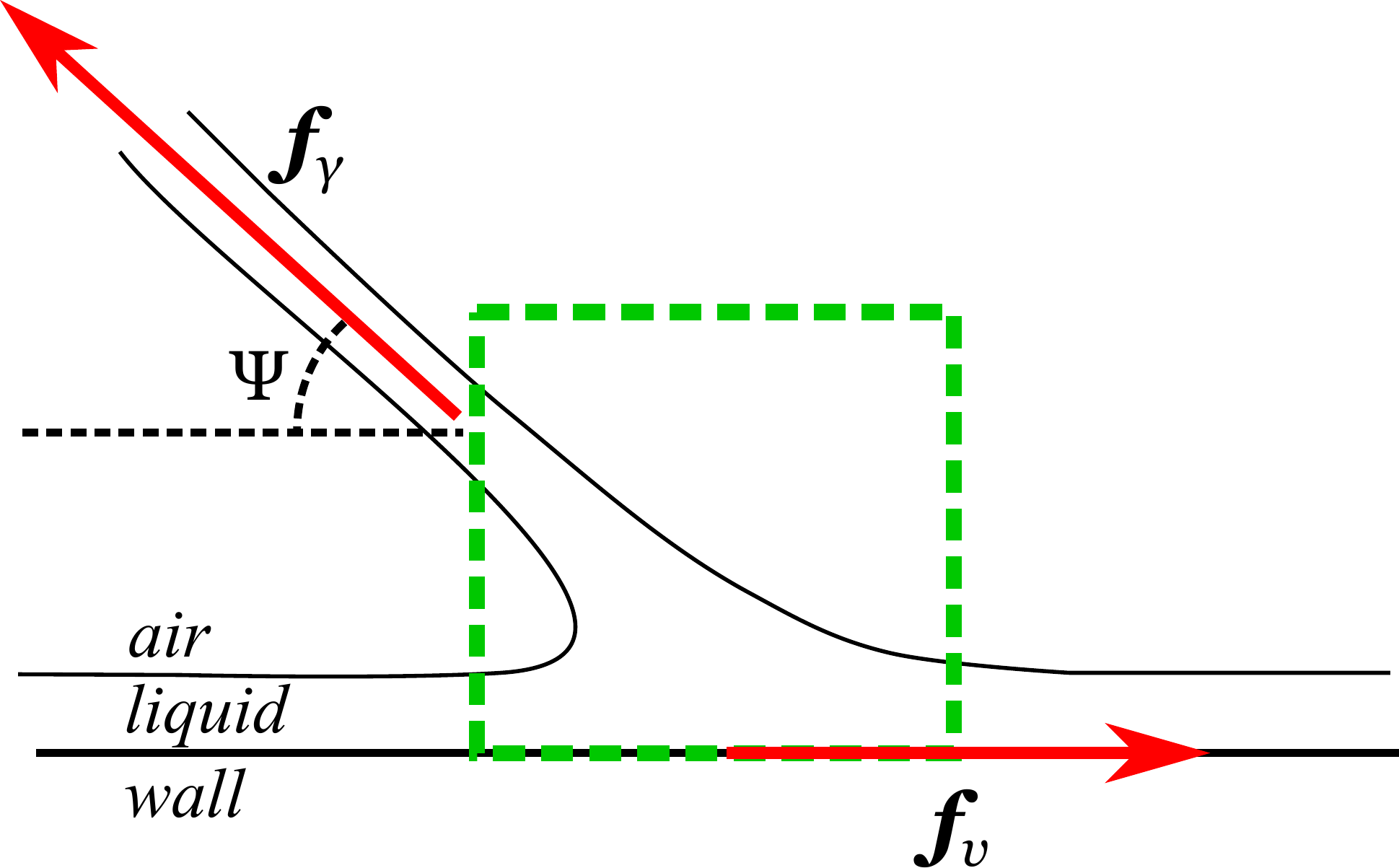}
  \caption{Momentum balance in a plane perpendicular to the surface Plateau border (region inside
  dashed green square) in contact with the wall.\label{fig:momentum}}
\end{figure}

In all of these problems, the boundaries of the soap film are rigid ``frames" that are
fixed in position or subjected to infinitesimal changes to induce the collapse.
In this paper, we study a distinct class of dynamics where part of the boundary of the film
moves as a consequence of the surface evolution.  In this context, the moving boundary problem
gives rise to a boundary singularity.  Specifically, we study a problem introduced briefly in
earlier work \cite{PRE} consisting of a catenoid film supported in the usual way by two circular
rings, but which is split into two hemicatenoids by a glass plate oriented perpendicularly to the
rings, as in Fig. \ref{fig1}.  The closed boundary of each half of the film thus consists of two
rigid, stationary semicircular frames and two curved, movable surface Plateau borders (SPBs) connecting 
the intersections of those frames with the glass plate.  As such, this is precisely of the
class of problems studied by Courant \cite{Courant_surfaces} as a generalization of the work
of Douglas \cite{Douglas1,Douglas2} on the existence of minimal surfaces bounded completely by 
Jordan curves.  It is one of a class of problems involving capillary surfaces
in contact with a wall, subject to various boundary conditions \cite{BostwickSteen,Akbari1,Akbari2}.
As shown in cross section in Fig. \ref{fig:momentum}, the lateral balance of forces that occurs 
in the surface Plateau borders involves the 
component of the force ${\bf f}_\gamma$ 
parallel
to the wall due to surface tension $\gamma$ 
and the frictional force ${\bf f}_v$ due to motion of the film at velocity $v$.  
In the case of a minimal surface, we know from Plateau's laws that the contact angle 
$\Psi=\pi/2$ and corresponds to equilibrium, i.e. $v=0$. 
The dependence of ${\bf f}_v$ on the sliding velocity $v$ is a well-studied problem in foam 
rheology \cite{Den2005Wal,Den2006Foa} and is strongly correlated with the nature of the 
gas-liquid interfaces, whose properties lie between the limiting cases of a
stress-free surface and an incompressible surface covered with surfactants \cite{Cantat2013Liquid}. 
Here, in contrast to the case of foams, it is the three-dimensional shape of the 
soap film that drives the surface Plateau border motion and thereby controls the dynamics.

The experimental setup in Fig. \ref{fig1} is related to the 
general problem of a triple line that becomes unstable. Examples include the collapse of liquid or 
air bridges during dewetting  \cite{reiter_dewetting_1992,reiter_unstable_1993} and air entrainment in 
dynamic wetting \cite{burley_experimental_1976,blake_maximum_1979,benkreira_air_2008}. These situations 
are characterized by small bubbles and droplets, respectively, that are left behind on the substrate. 
The instability of liquid bridges is also central to the detachment of droplets from 
moving drops 
\cite{podgorski_corners_2001,grand_shape_2005,engelnkemper_morphological_2016,liang_pearls_2017} or 
liquid bridges \cite{beltrame_rayleigh_2011,diez_instability_2012,singh_breakup_2017}. In each case, 
the dynamics is essentially that of a hemicatenoid in contact with a solid wall. 
For the case of soap films, the dynamics of a collapsing hemicatenoid is also relevant to 
bubble fragmentation inside foams confined in porous media \cite{Ger2016Flo,Ger2017FLa} or 
in microfluidic devices \cite{Lio2013Nei}.  

We study the collapse of hemicatenoids in the geometry of Fig. \ref{fig1}, and compare
the experimental measurements with a numerical model for the film evolution that incorporates the 
friction laws for surface Plateau border motion.
The model exploits the separation of time scales between the fast film motion and the slow creep of the SPBs.
Given this time-scale separation, the film can be approximated by a minimal surface that spans the support
rings and the Plateau borders, and its local contact angle with the plate determines the SPB speed.
This situation is similar to the retraction of a soap films inside an elastic ring \cite{Box2020Dyn}, 
whose shape and dynamics are linked to surface tension on one side as well as on the elastic and inertial properties of the ring on the other side. In that spirit, the properties of the ring are replaced here by the frictional properties of the surface Plateau border.

In Section \ref{expts} of this paper we summarize the experimental methods used both in the 
study of hemicatenoid collapse and also in the related problem of 
SPB reconnection on surfaces described in the Discussion. The main experimental
results are presented in Sec. \ref{results}, while the formulation of the
dynamical model used in numerical studies is given in Sec. \ref{model}.  Section \ref{numresults} 
presents results of those studies and a comparison with experiments.  The concluding Discussion
section \ref{discussion} connects these results to surface Plateau border reconnection and
other problems in foams.

\section{Experimental methods}
\label{expts}

As in previous work \cite{PRE}, soap solutions were obtained by dissolving tetradecyl trimethyl 
ammonium bromide (TTAB) into deionized water and 
adding glycerol in order to vary the viscosity.  
The concentration of TTAB was $3$ g/l for the aqueous solution containing 
no glycerol and was raised to $6$ g/l for the solutions containing glycerol to enhance the stability of the 
soap films. The viscosity $\eta$ varies from $1.0$ to $77$ mPa$\cdot$s over the range of glycerol 
concentrations, while the surface tension is nearly constant at $35-38$ mPa$\cdot$m
\cite{Cohen2014Inertial,Cohen2015Drop}.  Fluorescein was added to aid
visualization.

In the experiments on hemicatenoids, a full catenoid is first formed between two coaxial circular rings of 
radius $R=4\,$cm, whose distance apart is adjusted with a micrometric linear stage. A glass plate of 
width slightly less than $2R$ and thickness of $1\,$mm was introduced and passes through the 
diameters of both rings. In this way we create two independent hemicatenoids that become unstable 
once the distance $2d$ between the rings becomes larger than a critical value.

\begin{figure*}[t]
  \centering
  \includegraphics[clip=true,width=1.8\columnwidth]{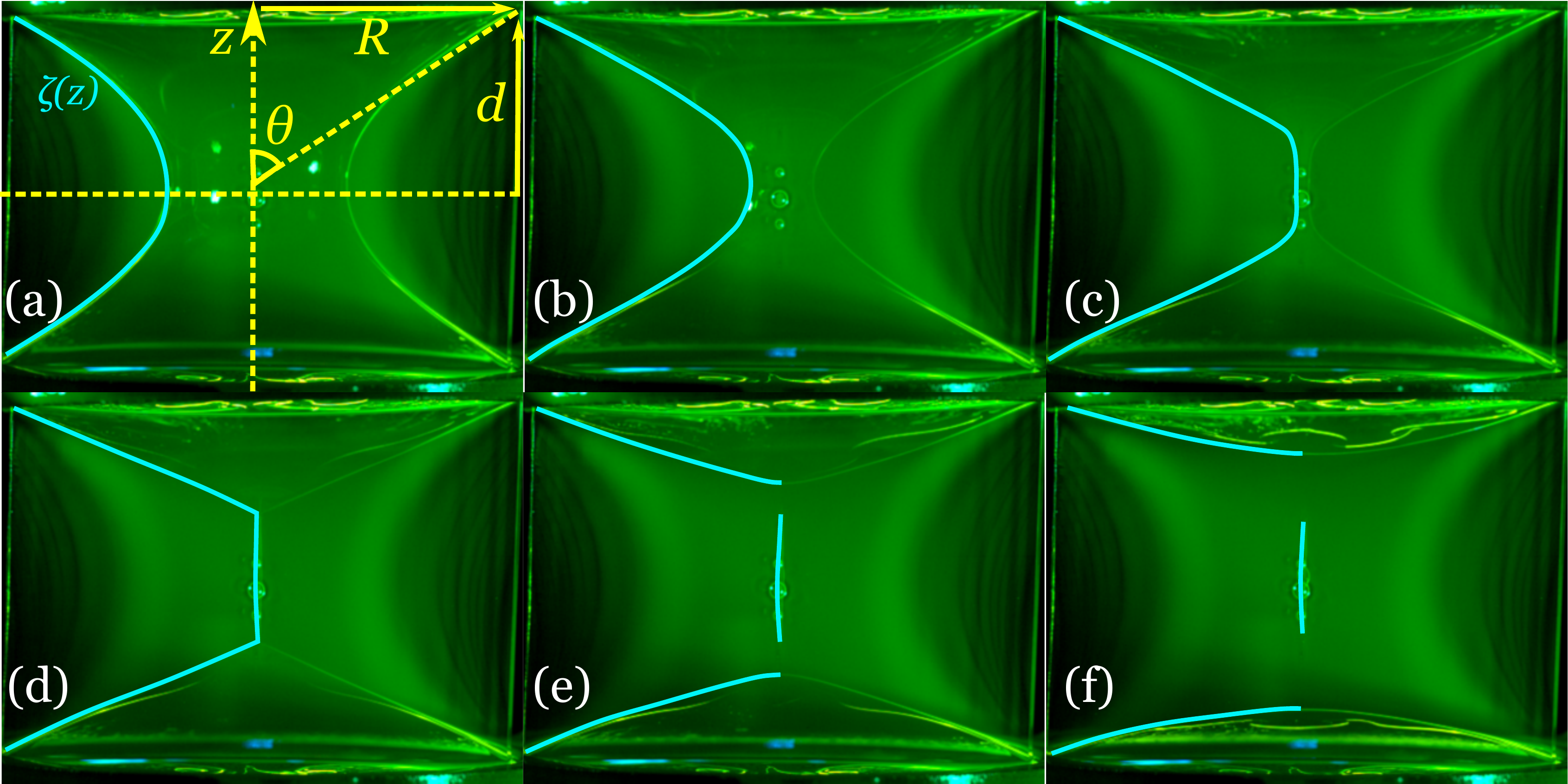}
\caption{Experimental collapse sequence of a hemicatenoid stretched beyond criticality.  Images are taken at 
times $t-t_p=$-51 ms, -16 ms, -6 ms, 0 ms, 10 ms, 35 ms.
Diagram indicates tangent angle $\theta$ of the interface at the upper wire frame.  Cyan contours are tracings of the
left-hand surface Plateau border of the hemicatenoid facing the camera.}
\label{fig3}
\end{figure*}

In separate experiments on SPB reconnection discussed in Sec.\,\ref{discussion}, 
we used a different soap film chemistry to slow
down the dynamics \cite{twist} for accurate visualization. We used deformable, 
transparent tubing with a large diameter ($6.4\,$mm) and a SLES/CAPB/SLES mixture \cite{Golemanov2008} 
known for its high dissipative properties \cite{Denkov2009}.  We mixed $6.6$\% of sodium lauryl ether sulfate 
(SLES) and $3.4$\% of cocamidopropylbetaine (CAPB) by weight in ultra pure water, and then dissolved $0.4$\% by 
weight of myristic acid (MAc), by stirring and heating at $60^\circ\,$C for one hour. We then diluted this solution 
by a factor of $20$ with water containing fluorescein at a concentration of 0.5 g/L. 

The dynamics of the collapsing soap films were recorded using a color high speed camera (Phantom V641, Ametek) at 
speeds up to $4,000$ frames per second while the films were illuminated from multiple directions with 
arrays of cyan LEDs.  The ESI\dag\, includes a video of the collapse of a 3d catenoid and videos of
collapsing hemicatenoids with $\eta=9.6$ and $77\,$mPa$\cdot$s.

\section{Experimental results}
\label{results}

As is well known, a full catenoid spanning two circular frames a distance $2d$ apart
becomes unstable when $d/R$ exceeds the critical value $0.663...$ \cite{PRE}.  
Using the setup shown schematically in Figure \ref{fig1}, we found 
experimentally that essentially the
same threshold exists in the case of a hemicatenoid.
In capturing images of this process with a high-speed camera aimed along the normal to the plate,
we naturally see two hemicatenoids, one in front and the other behind the plate.  
It is inevitable that they are slightly
different due to imperfections in the setup, and therefore they tend to collapse at slightly different times. 
We therefore introduced a deliberate, infinitesimal bias in the position of the plate in order to assure 
that the hemicatenoid
facing the camera would collapse last.  The consequence of this is seen in Fig. \ref{fig3}, where the bubbles
in the central region of the images are the satellite bubbles left over after the collapse of the hemicatenoid 
that was behind the plate. The cyan tracings in Fig. \ref{fig3} indicate one of the 
two surface Plateau borders 
of the hemicatenoid facing the camera.

The sequence of shapes of the SPB is strikingly similar to the equivalent 
sequence for three-dimensional
catenoids obtained by taking longitudinal cross-sections in the plane of symmetry \cite{PRE}.  We note in particular the
evolution toward a shape consisting of two conical films connected by a quasi-cylindrical region (panel (c)),
with a characteristic angle $\theta^*\simeq 67.5^\circ$ at the pinchoff time $t_p$ [panel (d)].  After pinchoff, the two
disjoint films still attached to the frame relax towards the two half discs spanning the wire loops,
while the remnants of the central cylinder slowly round up to form satellite bubbles, as seen in panel (f). 

In quantifying the observed film shapes, we label the SPB location as
$\zeta(z)$, and define \cite{PRE} the dimensionless radius $r=\zeta/a$, where $a=0.5524 R$ is the critical 
catenoid waist radius.
Similarly, we use the half disc area to define the dimensionless film area ${\cal A} = A/\pi R^2$, with
${\cal A}\simeq 1.199\ldots$ for the critical hemicatenoid and ${\cal A}=1$ for the configuration with two hemidiscs.
The area is calculated from the catenoid contour assuming axisymmetry. This requires that the 
contact angle on the glass is close to 90$^\circ$, as observed in experiments. While we know that this is not 
strictly fulfilled at all space-time points in the experiments, it is a convenient simplification for the measurements.
We have used the same assumption in the simulations to enable direct comparison of the two; the error in this approximation increases as the contact angle $\Psi$ decreases away from $90^\circ$ and pinchoff is approached.

 \begin{figure*}[t]
  \centering
 \includegraphics[clip=true,width=2.0\columnwidth]{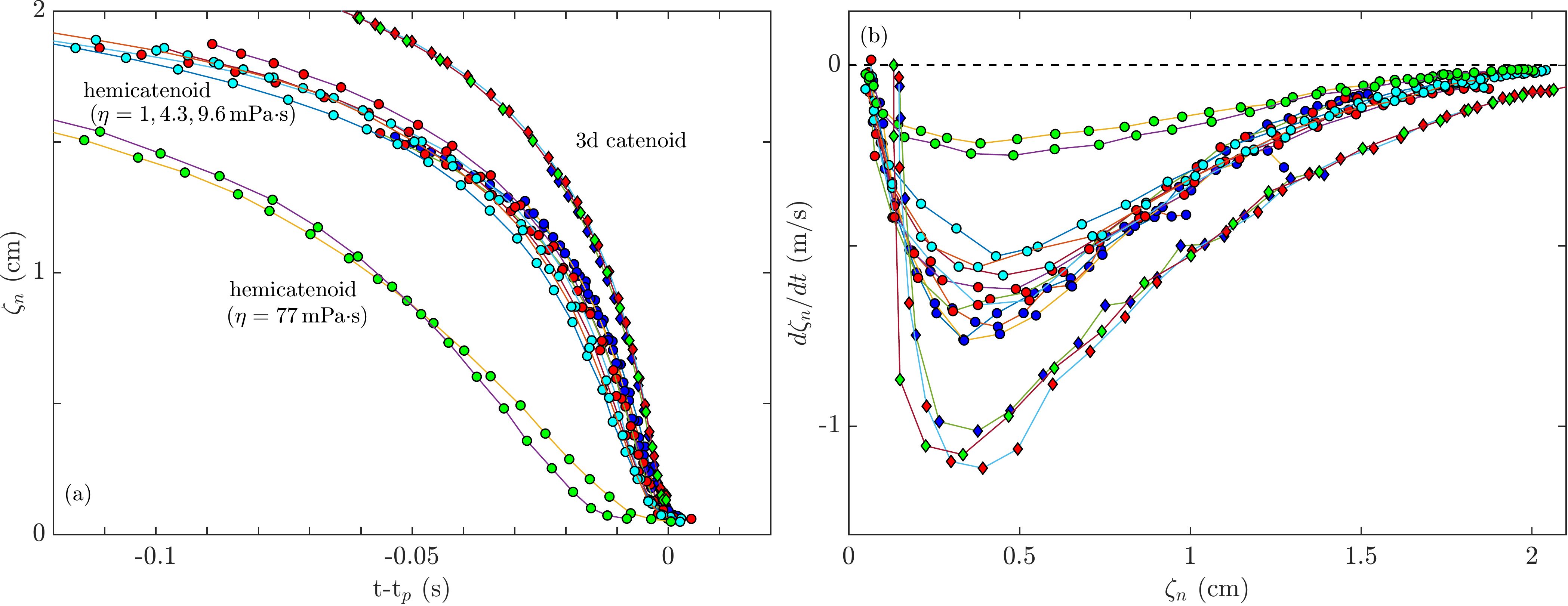}
  \caption{Time evolution of neck radius and contraction speed for three-dimensional catenoids
  and hemicatenoids.
  (a) The neck radius $\zeta_n$ as a function of time relative to pinch time 
  for the various cases as labelled.
  (b) Contraction speed versus neck radius.}
  \label{fig4}
\end{figure*}

From the measure SPB locations we determine the time evolution of ${\cal A}(t)$, the minimum neck radius 
$\zeta_n(t)$, and the 
neck contraction speed $d\zeta_n/dt$ for hemicatenoids and full catenoids (also with ring radius
$R=4\,$cm).  The data
for $\zeta_n$ shown in Fig. \ref{fig4}a illustrate how the contraction dynamics of even the least viscous
hemicatenoids is slower than the 3d case, and that the hemicatenoid dynamics progressively slows down
with increasing viscosity.  In contrast, the 3d data shows little if any dependence on viscosity, indicating
that the 3d balance of forces is between surface tension and air inertia.

Figure \ref{fig4}b plots the neck contraction speed as a function of
the neck radius.  From this we see that all the data sets display a maximum contraction speed at 
a neck radius $\zeta_n^*\approx 4\,$mm, independent of the film viscosity.  This is clearly associated with
the incipient formation of a central satellite droplet.  The lack of variation with $\eta$, and the
presence of the central plate, indicates
that it is a geometric effect in the sense that $\zeta_n^*$ should scale with the loop radius $R$.
This hypothesis is borne out by comparison with numerical studies of inviscid capillary breakup \cite{ChenSteen},
where the maximum contraction speed is reached for $\zeta_n/R\approx 0.1$.
While the radius of maximum contraction speed is common to both 3d catenoids and hemicatenoids,
the neck radius at which the speed vanishes for hemicatenoids is approximately one half of that for
the 3d case.  This difference arises from the very different local geometries of the
collapsing necks in the two cases.  Whereas the 3d catenoid maintains axisymmetry during collapse,
the hemicatenoids do not, as discussed in Section \ref{model} below.

We can understand the crossover between inertial and viscous effects in film collapse by means of a scaling argument
for characteristic film speeds $U$.
Under inertial dynamics, we expect the Keller-Miksis \cite{KellerMiksis} scaling $U\sim (\gamma R /\rho_a)^{1/2}$ to 
hold, where $\rho_a$ is the density of air, while in the viscous limit the speed $U\sim \gamma/\eta$ 
is associated with the capillary number of order unity \cite{Cantat2013Liquid}. The ratio of these velocities defines 
the Ohnesorge number of this problem $\textrm{Oh} = \eta/\sqrt{\gamma \rho_a R}$, where large (small) values 
correspond to frictional (inertial) dynamics, respectively. The critical viscosity $\eta_c$ for a balance
between the two effects is at $\textrm{Oh} =1$, namely $\eta_{c} \simeq 40\,$mPa$\cdot$s, which is consistent with 
the observations in Fig. \ref{fig4}.

\section{Model and simulations}
\label{model}

A moving meniscus in contact with a wall (an SPB) experiences a frictional force per unit length 
whose form depends on the spatial distribution of dissipation, which itself depends on 
the stress and physical chemistry at the liquid-air interface. This dissipation can be written as the 
sum of two terms \cite{Cantat2013Liquid}, both of which 
depend on the capillary number $Ca = \eta v/\gamma\,$.  The first term represents the
contribution from dissipation inside the wetting film, and scales as $Ca^{1/3}$,
while the second is the contribution from dissipation inside the SPB, and scales as $Ca^{2/3}$. 
The latter contribution is associated with mobile surfactants and stress-free interfaces.  Otherwise, 
both contributions are present and over a large range of $Ca$, 
this combined frictional law can be approximated as $f_v\sim Ca^n$ with $n$ in the range 
$1/2-2/3$. 
This empirical law 
holds for other systems in which the frictional force is
the dominant one in the balance with the capillary driving force, such as liquid and air bridges.

The model we study here rests on this assumption, namely that 
the capillary forces are balanced by the viscous friction at the contact between the soap film and 
the glass plate. That contact consists of an SPB 
\cite{Cantat_et_al_book} that connects the wetting film on the plate to the soap film. 
A slice through a plane perpendicular to this SPB (Fig.~\ref{fig:momentum}) shows that the 
driving force density is $f_\gamma \cos\Psi$, with $f_\gamma = 2 \gamma$.  
Balancing against the frictional force $f_v = A \gamma\, Ca^n$, with $A$ a dimensionless constant, 
we obtain 
\begin{equation}
    v = v_0 \left(\cos\Psi\right)^{1/n},
    \label{EqVelocity}
\end{equation}
where $v_0 = (\gamma/\eta)\left(2/A\right)^{1/n}$ is a characteristic speed.

\begin{figure}[hb]
  \centering
 \includegraphics[clip=true,width=1.0\columnwidth]{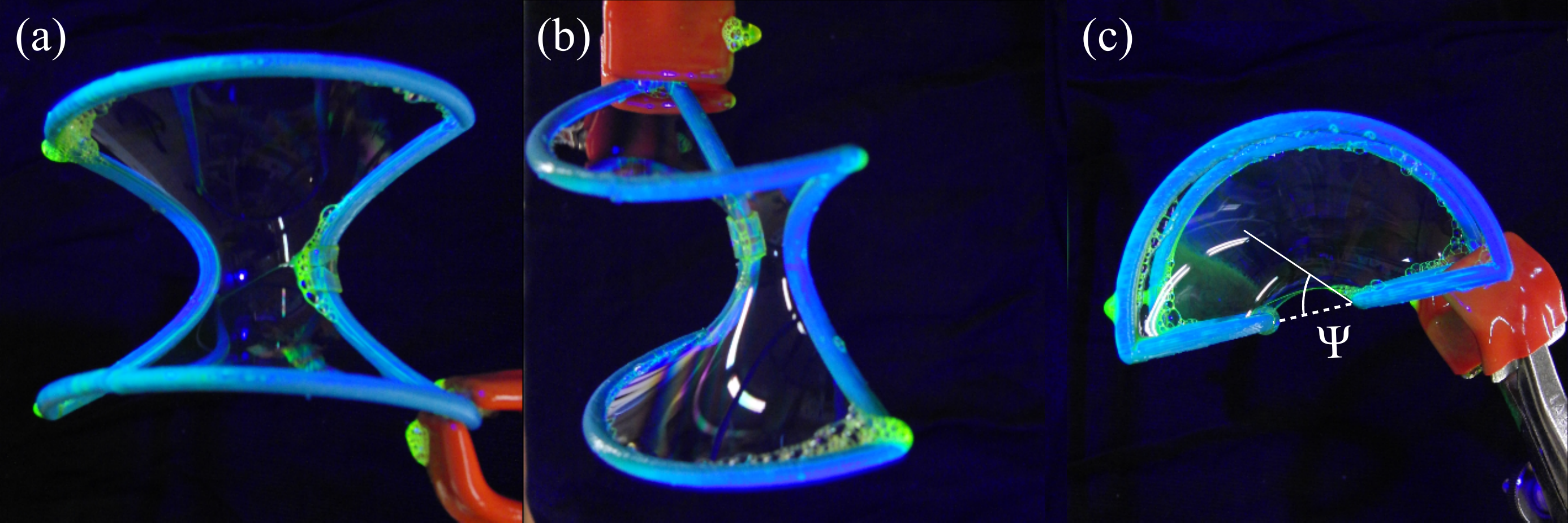}
  \caption{Soap films spanning a 3D printed frame consisting of two semicircles connected by a contour found 
  in numerical studies of film evolution.  Note in panel (c) how the minimal surface meets the
  narrowest part of the neck with a small angle $\Psi$.}
  \label{fig5}
\end{figure}

To apply this law to the evolution of hemicatenoids toward collapse, we exploit the separation
of time scales between the motion of the bulk surface, resisted only by the inertia of the
surrounding air, and that of the contact lines, and view the surface
spanning the moving contact lines and the two half-rings as in quasi-equilibrium. 
It is therefore a minimal surface at any instant of time, but does not necessarily meet
the glass plate with $\Psi=\pi/2$, as would be the case in static equilibrium in accordance with
Plateau's laws.  Instead, the contact line acts like a constraining wire, and the surface meets 
it at the angle that yields a minimal surface.  Figure \ref{fig5} shows an experimental 
realization of this concept, where 3D printed frames, consisting of two half-circular loops 
connected by contours obtained from the numerical studies below, support minimal surfaces
with contact angle $\Psi<\pi/2$.

\begin{figure*}[t]
  \centering
  \includegraphics[clip=true,width=1.6\columnwidth]{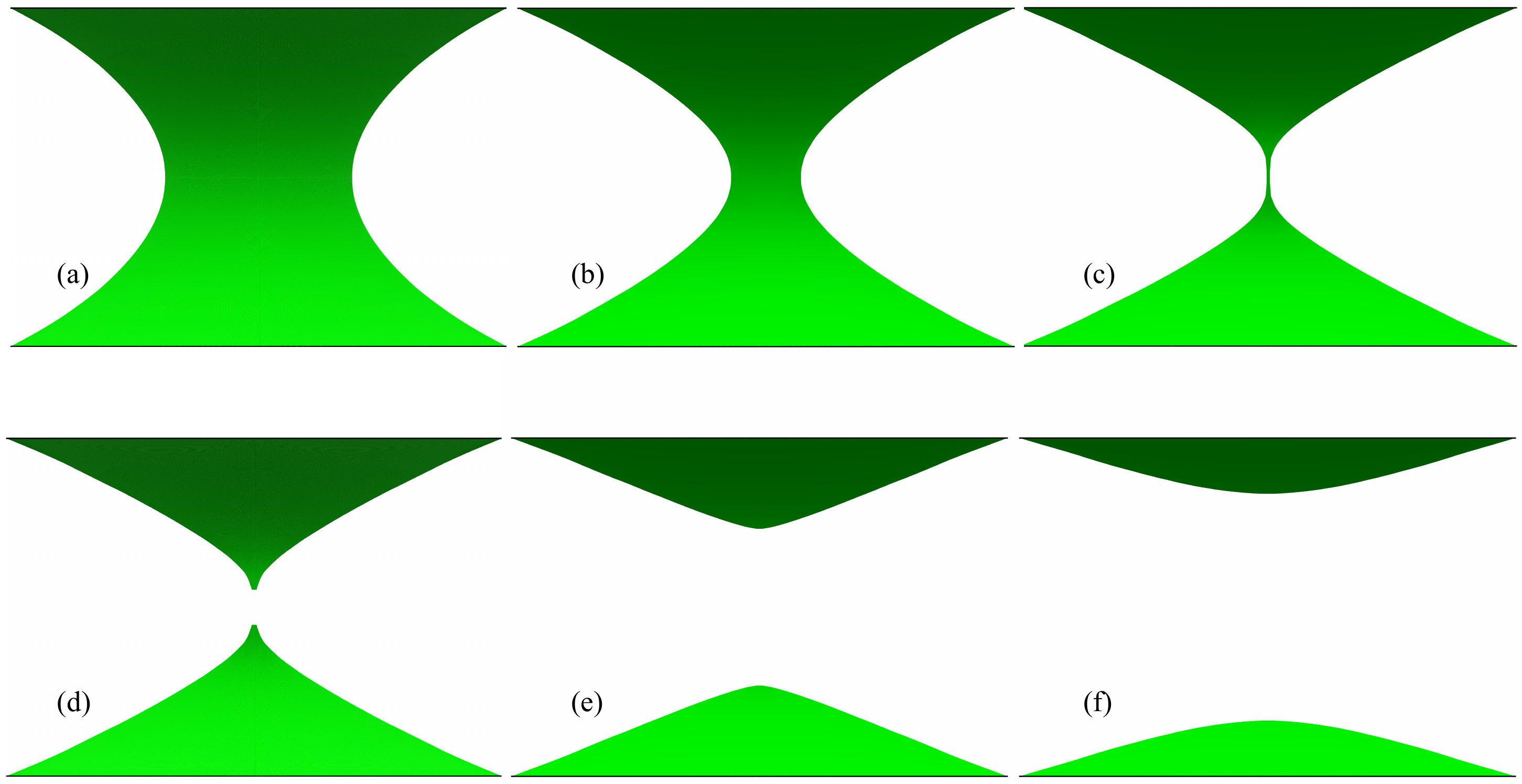}
  \caption{Image sequence in simulation ($n=2/3$, stretching factor 5\%), corresponding to the times at which the images in Figure \ref{fig3} were recorded.}
  \label{fig6}
\end{figure*}

The dynamics of evolving hemicatenoids obeying the laws specified above were obtained 
using Surface Evolver \cite{brakke92}.  We nondimensionalize lengths with the loop radius $R$ 
and scale speeds with $v_0$, leaving $R/v_0$
as a characteristic time scale in the motion.  Therefore,
changing $R$ in the initial conditions 
or  $v_0$ in \eqref{EqVelocity} corresponds to a dilation in time and all simulations can be 
rescaled with respect to time.  Hence, for a given power-law exponent $n$, there is only a single
simulation needed for each initial stretching factor.
The updating algorithm involves advecting each 
part of the SPB in the plane along its projected normal by a distance  $d\ell = v\,dt$,
with $v$ determined by
the local angle $\Psi$. The updated shape of the catenoid is obtained by finding
a minimal surface with the new position
of the contact lines, which then yields the updated $\Psi$ along the SPBs, and the 
velocity for the next time step.

By symmetry, we need only simulate one quarter of the hemicatenoid. We fix the plate thickness 
at $0.008R$ and first create a stable catenoid with $d = 0.65R$, close to the critical value
for three-dimensional catenoids, and a triangulation in which triangle edge lengths lie in the 
range $0.004-0.06$. 
To begin the evolution, we stretch the catenoid by a few percent  and fix all points along the SPB.
Motion then proceeds in time steps of $dt = 1\times10^{-3}$ using the algorithm outlined above,
where the relaxation to a minimal surface is done to $7$ significant figures.
The initial stretching is needed to match the initial area in experiments and in simulation. It corresponds 
to 5 and 7.2\% for the simulations of $\eta = 9.6$ and 77 mPa.s respectively. 

\begin{figure}[t]
  \centering
  \includegraphics[clip=true,width=0.75\columnwidth]{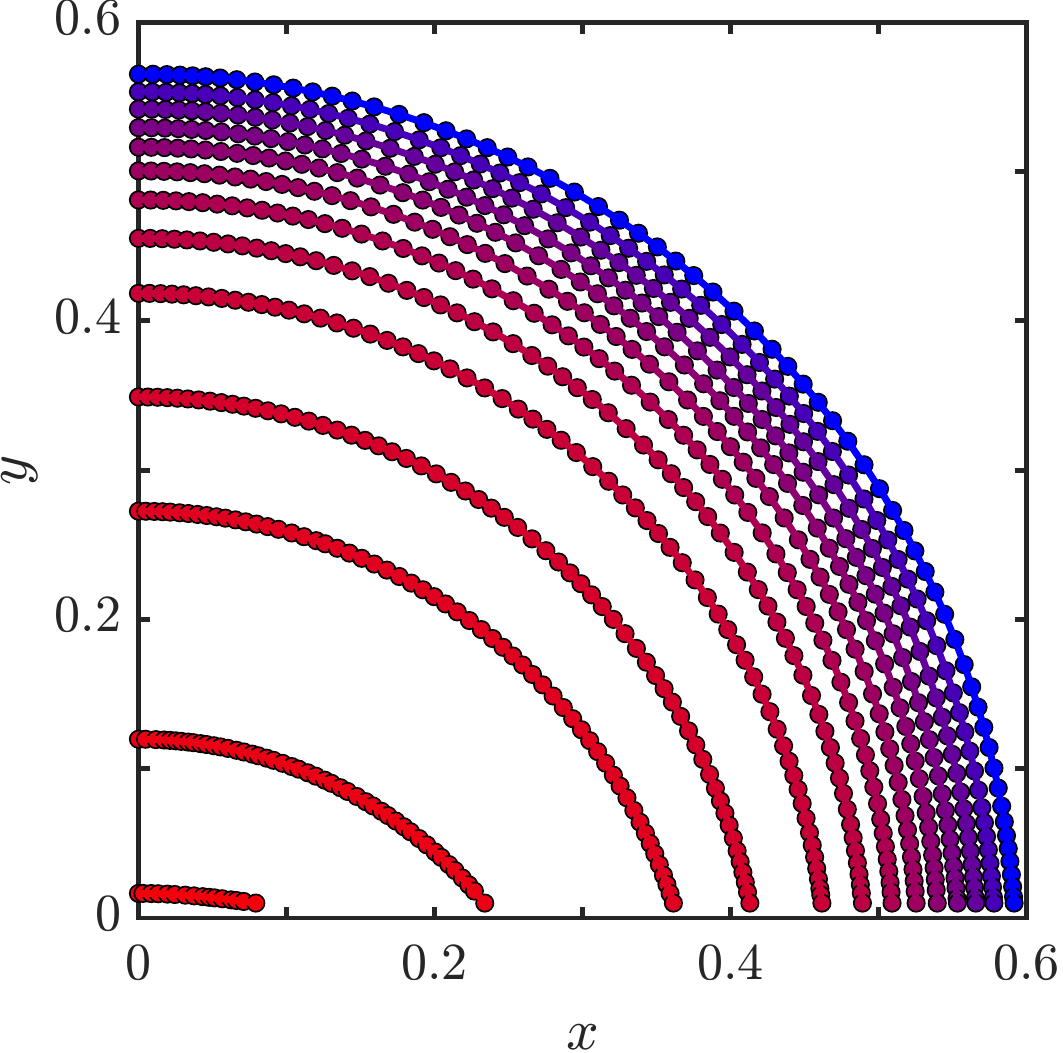}
  \caption{
Numerical results for cross-sectional profiles in the plane $z=0$ for $t<0$, color-coded from blue 
(early) to red (late).  The initially circular profile becomes extremely flattened, coincident 
with the contact line speed reaching its maximum.}
\label{fig7}
\end{figure}

In the numerics, several quantities are calculated as functions of time: the dimensionless area
$\mathcal{A}$, the dimensionless neck radius $r_n$ as defined in Sec. \ref{results}, 
and the tangent angle $\theta$ at the junction of the contact line and the supporting half-ring.
As in the presentation of the experimental results, we define the zero of time as the 
moment when the apparent neck radius vanishes and the topological transition occurs. In simulations, 
a cutoff radius of $0.02R$ is introduced to trigger the transition. This choice is not critical 
in the analysis since the radius goes to zero in a finite time and the transition is well defined, as 
seen in Fig.~\ref{fig7}.

\section{Comparison of Experimental and Numerical Results}
\label{numresults}

\begin{figure*}[t]
\centering
\includegraphics[clip=true,width=1.75\columnwidth]{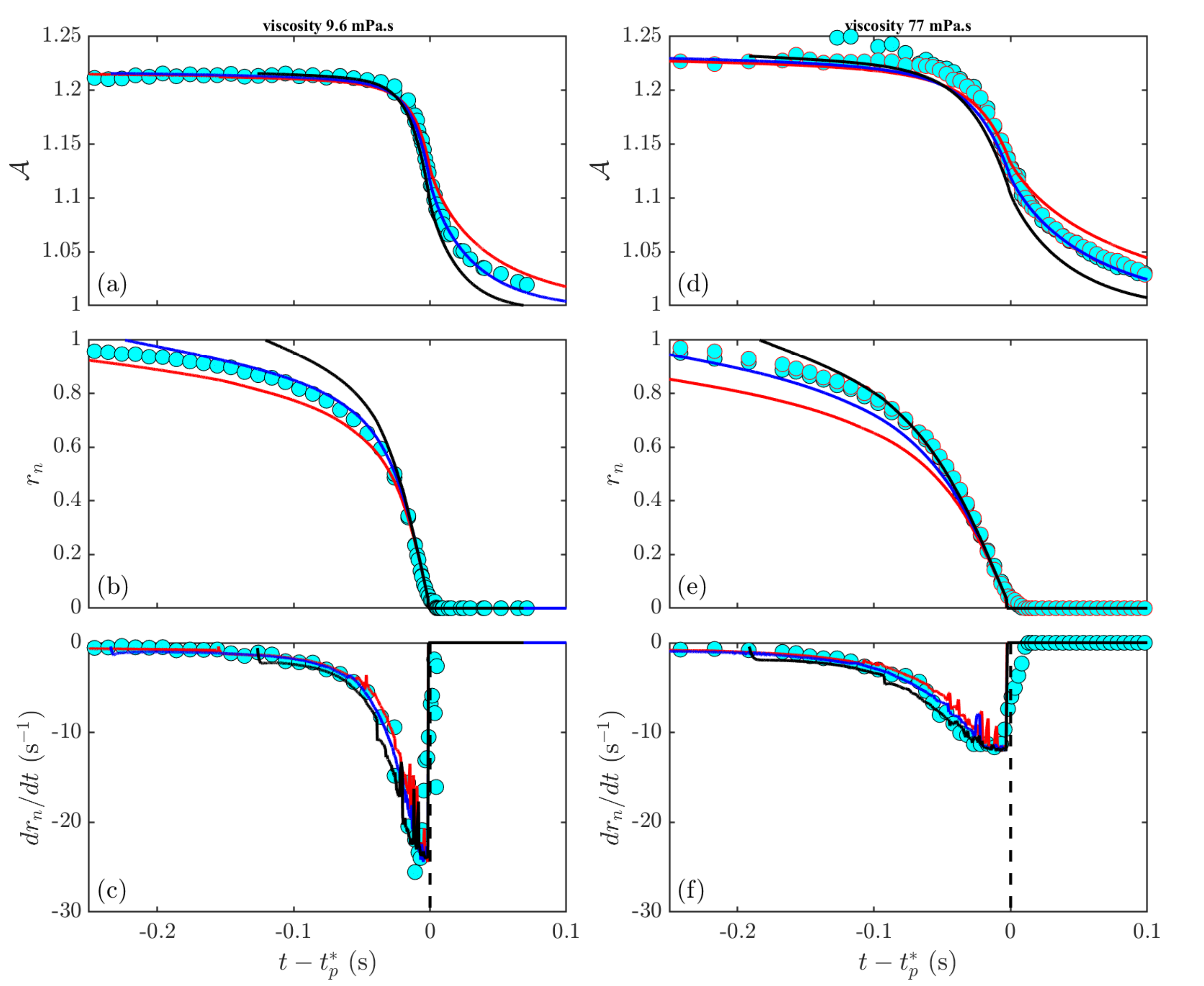}
  \caption{Rescaled area, neck radius and neck contraction rate as functions of time. Simulations were 
  performed with force-law exponent $n=1$ (black), $2/3$ (blue) and $1/2$ (red). Time has been
  scaled to match the maximum neck rate.}
  \label{fig8}
\end{figure*}

We now describe the key results obtained with the numerical implementation of the model
described in Section \ref{model}.  The full temporal evolution of a collapsing hemicatenoid
is shown in Fig. \ref{fig6} (and in a video in the ESI\dag), while Fig. \ref{fig7} shows the
cross-sectional shapes of the narrowest part of the neck.  We observe that the initial
cross-section is nearly circular, with the angle $\Psi\approx \pi/2$, as expected for a minimal
surface.  However, as the collapse sequence progresses, the neck becomes ever flatter and
$\Psi$ tends to zero. Within the assumptions of the model \eqref{EqVelocity}, this implies that the neck 
contraction speed $d\zeta_n/dt$ tends to the 
extremal value $-v_0$ before pinchoff.
As in collapsing three-dimensional catenoids \cite{PRE}, we observe that the film shape connecting the pinchoff region to the supporting wire loops is close to a (half) cone.  

The experimental data shown in Fig. \ref{fig4} for the neck radius and neck contraction speed, as well 
as the experimental film area are plotted in Fig. \ref{fig8} along with the numerical results obtained 
for three different values of the exponent $n$ of the friction law \eqref{EqVelocity}.
Simulations are rescaled with respect to experiments using the curves of the neck contraction speed 
as a function of time. The value of $v_0$ in simulations is set, for each viscosity $\eta$, 
by the maximum contraction speed observed in experiments. 
This sets the time axis up to an additive constant. There is a noticeable difference between experiments and
simulations close to pinchoff. In simulations, the neck contraction speed remains constant in the last 
instants before pinchoff at its extremal value $-v_0$. In experiments, the neck contraction speed reaches a 
maximum, used to infer $v_0$, and decreases thereafter to zero. As mentioned previously, this is associated 
with the complex dynamics 
close to pinchoff due to the air trapped in the neck, as observed in 3D catenoids. 
As a consequence, we define $t_p^*$ as the hypothetical 
pinchoff time assuming that the neck contraction speed maintains its extremal value (note that $t_p^* = t_p$ 
in simulations) and plot all the data as a function of $t-t_p^*$ in Fig. \ref{fig8}. This allows a 
quantitative comparison between experiments and simulations. For the two viscosities studied in Fig. 
\ref{fig8}, we see a clear tendency to favour the exponent $2/3$ over $1/2$ and $1$. This nonlinear friction law thus seems 
consistent not only with the presence of mobile surfactants and stress-free interfaces,
but also implies that dissipation occurs within the SPBs. 

\begin{table}[t]
\begin{center}
Experiments
\begin{tabular}{ccccc}
\hline
$\eta$ (mPa.s) & 1.0 & 4.3 & 9.6 & 77 \\
\hline
$\theta^*$ (deg) & $68.0\pm1.0$ & $68.0\pm1.0$ & $67.5\pm1.0$ & $66.6\pm1.0$  \\
\hline
\end{tabular}
Simulations\\
\begin{tabular}{ccccc}
\hline
$n$ & 1/2 & 2/3 & 1  \\
\hline
$\theta^* (deg)$ &  $65.8\pm0.6$ & $68.2\pm0.7$ & $72.1\pm1.1$  \\
\hline
\end{tabular} 
\caption{\label{tab:tableuno}
Apex angle of the Martini-glass configuration in simulation and experiment.
}
\end{center}
\end{table}

\begin{figure*}[t]
  \centering
  \includegraphics[clip=true,width=1.5\columnwidth]{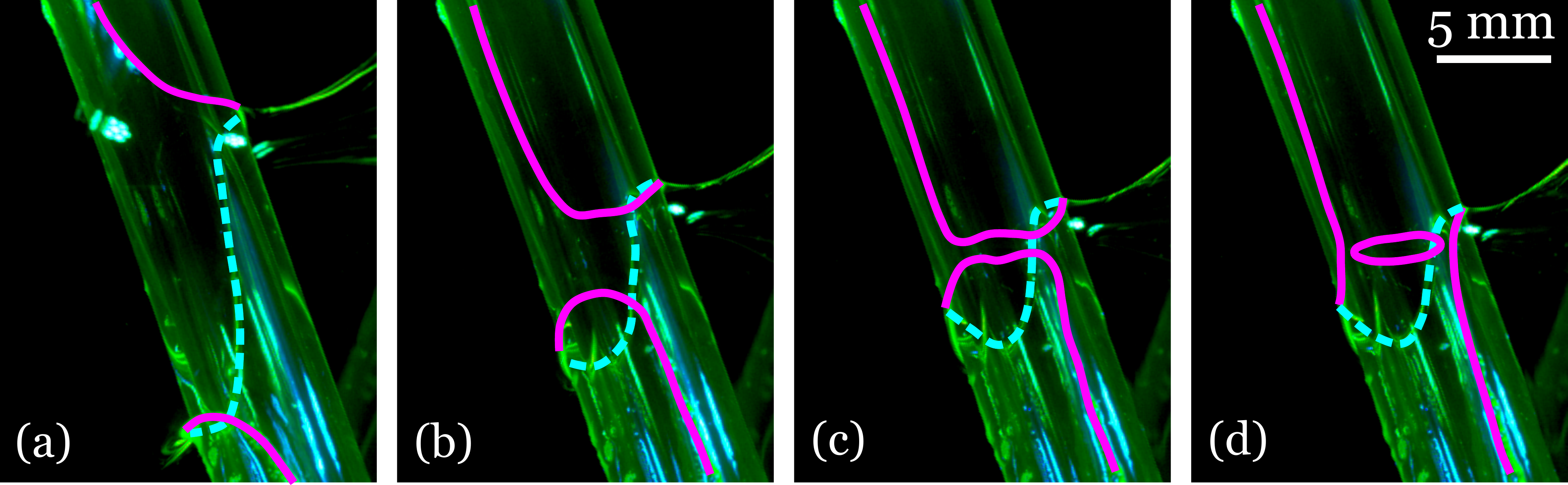}
  \caption{Surface Plateau border motion during a topological transition. Image sequence of the 
  final stages of SPB reconnection as a soap film M{\"o}bius strip transitions to a disc.
Foreground and background segments of the SPB are displayed by solid magenta and dashed cyan curves
respectively. Frames (a)-(c) are at times $-31\,$ms, $-12.5\,$ms,
and $-5\,$ms relative to the reconnection event in (d).}
  \label{fig9}
\end{figure*}

The value of the prefactor $A$ in the frictional force law \eqref{EqVelocity} 
can be inferred from the experimental measurements of $v_0$, using the exponent $n=2/3$ 
that is most consistent with the data. For the data of Fig. \ref{fig8}, we find 
for $\eta = 9.6$ and $77$ mPa$\cdot$s that $v_0 = 0.52$ and 
$0.26\,$m$\cdot$s$^{-1}$, respectively. These yield $A\approx 7.3$
and $2.9$, respectively, in agreement with the prediction $A\sim 5-7$ from lubrication theory 
in a simpler geometry \cite{Cantat2013Liquid}. For the two other viscosities 
($\eta = 1.0$ and $4.3\,$mPa$\cdot$s), the dynamics can not be described with the same approach 
since both viscous and inertial effects are present.

\addtocounter{footnote}{2}

In previous work \cite{PRE}, we have referred to the conical film shape as the ``Martini glass" 
configuration, and detailed a method to determine its appearance based on the shape of the film 
near the supporting ring. The method involves finding, within the time sequence, that contour 
whose shape displays the minimum standard deviation in a fit to a straight line. The critical 
vertex angle is then 
$\theta^*=\pi/2-\arctan(m)$, where $m$ is the associated slope of the line. For the case of 
3D catenoids, whose dynamics are driven by inertia, we found $\theta^*=68^\circ$. For hemicatenoids bounded 
by a solid wall, our measurements are reported in Table \ref{tab:tableuno}.
In experiment, this configuration is observed when the neck pinches. The stated uncertainties 
are estimated either over repeated experiments or by assuming a 10\% allowance on the standard deviation 
to find its minimum value. Both estimates give typical values around $1^\circ$. 
In the simulations, this configuration occurs slightly after the pinchoff time and for larger $n$ 
this configuration occurs later, and yields a larger $\theta^*$. 
\footnote[3]{For $n=1$ the local minimum is 
less pronounced.  We have tested a simulation with $n=4/3$ (data not shown). In this case, the 
local minimum 
does not exist as emphasized by the absence of a curvature inversion in the corresponding 
movie/image sequence 
(data not shown).}
In the simulations, the error bars are determined with the same 10\% criterion. 

We find that $\theta^*$ is almost constant in experiments and very close to what was observed  
experimentally with 3D catenoids \cite{PRE}, and which was found to be a consequence of the
fact that the first few active modes dominate the dynamics near the rigid frames. 
A slight decrease could be observed as the viscosity increases, but it is difficult to give strong statements 
given the error bars. We can notice that the value of $\theta^*$ obtained with $\eta = 9.6$ mPa.s is 
consistent with a 2/3 exponent. For $\eta = 77$ mPa.s, the value lies between what is obtained with 
exponents $1/2$ and $2/3$ in simulation, but this method to discuss the exponent is not as robust as 
the comparison through the whole time sequence as presented in Figure \ref{fig8} and is more sensitive 
to the exact and complex dynamics around the pinchoff time.

\section{Discussion}
\label{discussion}

In this work, we have studied the dynamics of surface Plateau borders that are driven by unstable soap films.
Unlike in the case of collapsing three-dimensional catenoidal films, the motion of hemicatenoids bounded by
solid walls is dominated by dissipation localized in the moving SPBs.
Whereas the film viscosity plays essentially no observable role in the collapse of 3d catenoids, we observe a strong
dependence on viscosity of the collapse dynamics of hemicatenoids.
Through comparisons between experimental observations and numerical simulations of a model that incorporates 
the widely used nonlinear friction law $f\sim Ca^n$ for the viscous force $f$ as a function of capillary number $Ca$, we found 
consistency with the exponent $n=2/3$ associated with mobile surfactants and stress-free interfaces.  
Despite this effect on the dynamics, the large-scale shapes of the collapsing films are insensitive to the film
viscosity, a fact that highlights the crucial role played by intrinsic geometric features. 

Surface Plateau border motion is also present in certain exotic topological transitions involving
soap films.  For example, in previous work\cite{twist,boundary,systolic} we studied the interconversion 
of a soap-film M{\"o}bius strip 
to a two-sided surface, and showed that the singularity associated with that topological transition
occurs at the boundary of the film and involves reconnection of the SPB.
Figure \ref{fig9} shows a close-up of the region of the incipient singularity, in which we see
that the film evolves in such a way as to make the SPB twist ever more tightly around the tube,
until two sections of the film touch each other and reconnect.
While this is similar to the 
neck of a half catenoid, which collapses with the formation of a satellite droplet,  
the detailed
dynamics of the twist-driven reconnection event and associated change in the orientation of the SPB 
in Fig. \ref{fig9} remain to be understood.

The dynamics that we describe here, driven by the shape of the interface and the motion of the 
associated SPB, is found more widely during the motion of surfactant-laden interfaces. In flow 
through porous media, for example during foam enhanced oil recovery~\cite{Cantat_et_al_book}, 
lamellae are forced under a pressure gradient through narrow pore throats between solid surfaces. 
The collision of two SPBs leads to processes such as pinch-off~\cite{Lio2013Nei}, in which two 
interfaces separate in the way that the hemicatenoid does, snap-off and 
lamella-division~\cite{Ger2017FLa}. These process lead to variations in bubble size, which 
then requires an adjustment to the pressure gradient required to mobilize the foam. SPB 
friction therefore controls the rate at which foam is generated in porous media.

\section*{Author Contributions}
All authors designed and performed the research, analyzed the results and wrote the paper.

\section*{Conflicts of interest}
There are no conflicts to declare.

\section*{Acknowledgements}
We are grateful to K. Brakke for provision and support of the Surface Evolver software and 
to M\'ed\'eric Argentina and Keith Moffatt for enlightening discussions.
This work was supported in part by the National Research Agency (ANR-20-CE30-0019) and by the 
French government, through the UCAJEDI Investments in the Future project of the National Research 
Agency (ANR-15-IDEX-01) (CR),
with additional support from the 
Engineering and Physical Sciences Research Council (EP/N002326/1 to SC; Established Career 
Fellowship EP/M017982/1 to REG \& AIP), 
from the Schlumberger Chair Fund (REG), and the European Space Agency (4000129502) (SC).

\balance

\bibliographystyle{rsc}

\end{document}